\documentclass[aps,twocolumn,showpacs,preprintnumbers,floats,amsmath]{revtex4}

\usepackage{graphicx}
\usepackage{dcolumn}
\usepackage{bm}
\usepackage{epsfig}
\begin{document}

\title{
Deconfinement phase transition in hybrid neutron stars from the
Brueckner theory with three-body forces and a quark model with
chiral mass scaling}
\author{G.~X.~Peng}
\affiliation{
 China Center of Advanced Science and Technology
 (World Laboratory), P.O.Box 8730, Beijing 100080, China\\
 Theoretic Physics Center for Science Facilities,
 Institute of High Energy Physics, CAS, Beijing 100049, China\\
 Department of Physics, Graduate University,
 Chinese Academy of Sciences, Yuquan Rd.\ 19A, Beijing 100049, China
         }
\author{A.~Li}
\affiliation{
 Dipartimento di Fisica e Astronomia, Universit\'a di Catania,
   viale Andrea Doria 6, 95125 Catania, Italy\\
 Department of Physics and Institute of Theoretical Physics and Astrophysics,
 Xiamen University, Xiamen 361005, China
         }
\author{U.~Lombardo}
\affiliation{
 Laboratori Nazionali del Sud, Istituto Nazionale di Fisica Nucleare,
 Via S.\ Sofia 62, 95123 Catania, Italy\\
 Dipartimento di Fisica e Astronomia, Universit\'a di Catania,
   viale Andrea Doria 6, 95125 Catania, Italy
         }

\date{\today}

\begin{abstract}
We study the properties of strange quark matter in equilibrium with
normal nuclear matter. Instead of using the conventional bag model
in quark sector, we achieve the confinement by a density-dependent
quark mass derived from in-medium chiral condensates. In nuclear
matter, we adopt the equation of state from the
Brueckner-Bethe-Goldstone approach with three-body forces. It is
found that the mixed phase can occur, for a reasonable confinement
parameter, near the normal nuclear saturation density, and goes over into
pure quark matter at about 5 times the saturation. The onset of
mixed and quark phases is compatible with the observed class of
low-mass neutron stars, but it hinders the occurrence of kaon
condensation.
\end{abstract}


\maketitle

\section{Introduction}

By far the study of neutron stars has been mainly focused on the
relationship between the equation of state (EoS) of nuclear matter
and the observed maximum mass. The connection has been achieved by
solving the hydrostatic equilibrium equations based on general
relativity. The first generation observed masses exhibited an
average value around 1.5 solar masses. This value requires a soft
EoS that can be easily obtained by introducing new degrees of
freedom like hyperons, kaons or quarks accompanied or not by a phase
transition. Sometimes the softening was so large that the neutron
star is predicted to collapse into a black hole, as for the SN1987A
\cite{BB}. In the new generation of observations the masses are
distributed within a large range, up to 2 solar masses, that
requires a stiff EoS, i.e., hadronic matter without new degrees of
freedom. Since it is hard to imagine pure hadronic matter to sustain
the high pressure predicted in the inner core, new scenarios have to
be advanced to explain the coexistence, in the phenomenology of
neutron stars, of low and high mass spectra.

Recently \cite{HAE} it has been argued that the two observed classes
of neutron stars might correspond to two different evolutionary
scenarios of neutron stars. In one case, the hot and dense remnant
of the supernova explosion rapidly evolves into a hybrid star, where
the transition to a quark phase softens the nuclear matter so that
$M\approx 1.5 M_\odot$; in the other case a slow evolution could lead the
neutron star to a large mass via a mass accreting from the coupling
with a white dwarf. From this point of view the destiny of the
remnant is strongly affected by the initial conditions, i.e. density,
temperature, leptonization degree etc. For instance, if the mass of
the remnant is below the mass threshold for quark nucleation the
transition to the quark phase is forbidden \cite{bombaci}. If the
mass is slowly accreting the transition is allowed. The role of
temperature or other parameters defining the initial state of a new
born neutron star has not yet been studied.

To investigate the possible phase transition to quark matter in
neutron stars, we need also to know the EoS of quark matter.
Although we have in hand the fundamental theory of strong interactions,
i.e., Quantum Chromodynamics (QCD), we still do not know the
true ground state. It is now generally expected that quark matter
is in the color-flavor locked phase (CFL) \cite{Alford1999} at extremely
high densities when the finite current mass of strange quarks becomes
unimportant. In the density range from nuclear saturation to CFL, there
may exist a rich and varied landscape of phases, e.g., the 2SC, g2SC,
gCFL etc. Presently, however, these phases suffer from the
so-called chromomagnetic instability problem for both the
two- \cite{HuangM2004} and three-flavor
\cite{Casalbuoni2005,Alford2005,Fukushima2005} cases. On the other
hand, experiments show that quarks become asymptotically free
rather slowly \cite{LiTD2005}. Therefore, in the present study
we are dealing with the ordinary strange quark matter (SQM)
\cite{WittenPRD,Jaffe}.

The special problem in studying the EoS of ordinary
quark matter is to treat quark confinement in a proper way. In the
conventional standard approach, an extra constant term, the
famous bag constant $B$, is added to the energy density of the system, which
provides a negative pressure to confine quarks within a finite
volume, usually called a `bag'. The quark mass is infinitely large
outside the bag, and a finite constant within the bag. A vast
quantity of investigation have been performed within the framework
of the bag model \cite{WeinerIJMPE15}.

As is well known, however, particle masses vary with environment.
Such masses are usually called effective masses. Effective masses
of hadrons and quarks have been extensively discussed, e.g.,
within the Nambu-Jona-Lasinio model \cite{Buballa99plb}
and within a quasi-particle model \cite{SchertlerNPA616}.
In principle, the density dependence of quark masses should be
connected to the in-medium chiral condensates
\cite{LombardoPRC72,PenggxNPA747}.

Taking advantage of the density dependence, one can describe
quark confinement without using the bag constant.
Instead, the quark confinement is achieved by the density
dependence of the quark masses derived from in-medium chiral
condensates \cite{PenggxPRC61,PenggxPRC62,WenxjPRC72}.
The two most important aspects in this model are the quark mass
scaling~\cite{PenggxPRC61,WenxjPRC72} and the thermodynamic
treatment~\cite{PenggxPRC62,WenxjPRC72}.
Both aspects will be reviewed in this paper.

In the present contribution, the transition from hadron phase (HP) to
strange quark phase (SQP) in the inner core of a neutron star is
investigated within the fully consistent nuclear and quark models.
In the hadron sector we adopt the equation of state from
Brueckner-Bethe-Goldstone (BBG) approach with three-body forces (TBF)
\cite{grange:1989,Zuowei02prc,shq1998prl}.
This theory, being a completely microscopic approach, can easily incorporate
degrees of freedom such as nucleon resonances [$\Delta$(1232) or
$N^*$(1440)], which are expected to appear at higher hadron densities.
It is found that the mixed hadron-quark phase can occur, for
reasonable values of the confinement parameter, a little above the
normal saturation density, and can undergo the transition  to pure
quark matter at about 5-6 times the saturation. This result is quite
different from the previous results from Nambu-Jona-Lasinio (NJL)
model in which the mixed quark phase can not appear at neutron-star
densities~\cite{Schertler99prc,njl}. Afterwards, the influence of
the mixed and quark phases on the structure of compact stars is discussed
by solving the Tolman-Oppenheimer-Volkov (TOV) equation and extracting
the mass-radius plots for neutron stars. Finally, it is shown that
the transition to the deconfined phase turns out to be incompatible
with the onset of kaon condensation.

\section{EoS of quark matter}

SQM has been one of the hot topics in nuclear physics since Witten's
famous stability conjecture \cite{WittenPRD}. In many
studies, the quark confinement was treated adopting the bag mechanism
\cite{Jaffe,Madsen}. An alternative approach to obtain
confinement is based on the density dependence of quark masses
\cite{FowlerZPC9}. This mechanism has been extensively applied to
investigating the properties of SQM
\cite{ChakrabartyPLB229,BenvenutoPRD51,Peng99prc,ZhangY02prc}. In
this section, we first give a short review of the two most important
aspects, with the main inconsistency of the original model. Then we present a
fully self-consistent thermodynamic treatment. The properties of SQM
will be given in the new treatment. In the present paper, however,
the main application of the new approach is the study of the phase
transition in compact stars.

\subsection{Confinement by density-dependent masses}

As mentioned above, the quark confinement in this
model is achieved by the density dependence of quark masses.
Therefore, the first important question is how to determine the
quark mass scaling which can reasonably produce confinement.
Originally, the interaction part of the quark masses was assumed
to be inversely proportional to the density
 \cite{FowlerZPC9,ChakrabartyPLB229}.
This linear scaling has been extensively applied to studying the
properties of SQM \cite{ChakrabartyPLB229,BenvenutoPRD51,%
Peng99prc,ZhangY02prc}.
There are also other mass scalings \cite{WangP00prc,PengGX97prc}.
Their main drawback is that they are pure parametrizations without
any convincing derivation. Therefore, a cubic scaling was
derived based on the in-medium chiral condensates and linear confinement
at both zero \cite{PenggxPRC61} and finite temperature \cite{WenxjPRC72}.
This new scaling has been applied to investigate the viscosity of SQM
and the damping time scale that is due to the coupling of the viscosity
and r mode \cite{Zheng04}, the quark-diquark equation of state
and compact star structure \cite{Lugones03}, the properties of strangelets
versus the electric charge and strangeness \cite{Wenxj07jpg}, and
the new solutions for CFL slets \cite{Wenxj07ijmpa}.
In the present paper, we use the chirally determined quark mass
scaling~\cite{PenggxPRC61,WenxjPRC72} to study the phase transition
in neutron stars. For this we need a completely self-consistent
thermodynamic treatment of the EOS of quark matter.

The thermodynamic treatment of the system with confinement via
the density dependent quark masses has been a long story.
Originally, the thermodynamic formulism was regarded the same
as the constant-mass case \cite{ChakrabartyPLB229}. In
this first treatment, the internal pressure can not be zero, and the
properties of SQM were rather different from those in the bag model.
But it was later pointed out that the difference was caused by the
incorrect thermodynamic treatment~\cite{BenvenutoPRD51}. It was found
that an additional term is to be added to the pressure and energy
expressions \cite{BenvenutoPRD51}. This second treatment makes it
possible SQM to be self-bound. However, two serious problems
came out: one is the unreasonable vacuum limits, the other one is the
discrepancy between the energy minimum and zero pressure.
It was shown that the added term in the pressure,
due to the density dependence of quark masses, should not be appended
to the energy. After discarding this term in the energy, while keeping
it in the pressure, the two inconsistencies mentioned above were
immediately removed \cite{PenggxPRC62}.
This third treatment has recently been
extended to finite temperature \cite{WenxjPRC72}. The thermodynamic
formulism in Ref.~\cite{PenggxPRC62} was also adopted in
Ref.~\cite{ZhangY02prc}, though a different quark mass scaling was
used there.

A common feature of the last two thermodynamic
treatments~\cite{BenvenutoPRD51,PenggxPRC62,WenxjPRC72},
as well as other recent references using this model \cite{ZhangY02prc},
is that they all regard the thermodynamic potential
the same as in the Fermi gas model. Because of the
additional term, the pressure becomes obviously not equal to the minus
thermodynamic potential density, contradicting the thermodynamic
equality $P=-\Omega$ for a homogeneous system.
One can also easily check that the fundamental differentiation
equality $\mbox{d}E=\sum_i\mu_i\mbox{d}n_i$ for homogeneous systems at zero
temperature was not fulfilled in the mentioned references.

In the rest of this section, we will present a
fully self-consistent thermodynamic treatment of
the confinement by density-dependent mass model (CDDM).

\subsection{Self-consistent thermodynamics in CDDM}
\label{thermofm}

Let us consider a quark model with three flavors. Denoting the Fermi
momentum in the phase space by $\nu_i$, the particle number densities
can then be expressed as
\begin{equation} \label{nimod}
n_i
=g_i\int \frac{\mathrm{d}^3{\bf p}}{(2\pi\hbar)^3}
=\frac{g_i}{2\pi^2} \int_0^{\nu_i}\, p^2\,\mbox{d}p
=\frac{g_i\nu_i^3}{6\pi^2},
\end{equation}
and the energy density as
\begin{equation} \label{Emod}
E =\sum_i\frac{g_i}{2\pi^2}\int_0^{|\nu_i|}\sqrt{p^2+m_i^2}\,p^2\,\mbox{d}p.
\end{equation}

Equations (\ref{nimod}) and (\ref{Emod}) are familiar expressions,
where the summation index goes over all considered particle types.
To let the model be valid for both particles and anti-particles, the
particle number density, or accordingly, the Fermi momentum, are
formally assumed to be negative for anti-particles. Therefore, in the
upper limit of the integration, the absolute value has to be taken.

If the particle masses $m_i$ are constant,
the relation between the Fermi momenta $\nu_i$ and the
chemical potentials $\mu_i$ is
\begin{equation} \label{nuimuifree}
\nu_i=\sqrt{\mu_i^2-m_i^2}\
\ \mbox{or}\ \
\mu_i=\sqrt{\nu_i^2+m_i^2}.
\end{equation}

As is well-known, however, the quark mass depends on
density and temperature. In principle, the quark mass scaling
should be determined from QCD, which is obviously
impossible presently. Based on the in-medium chiral condensates,
a cubic scaling law was derived at zero temperature \cite{PenggxPRC61},
and it has been recently extended to finite temperature \cite{WenxjPRC72}.
At zero temperature, we have the simple cubic scaling
\begin{equation}
m_q =m_{q0}+\frac{D}{n^z},
\label{mqT0}
\end{equation}
where $m_{q0}$ is the quark current mass, $n$ is the total baryon number
density, the exponent of density is $z = 1/3$ \cite{PenggxPRC61}, and the
constant $D$ is to be discussed a bit later.

In the following, we show that the density-dependence of
particle masses will modify the Fermi momentum, i.e.\
the relation in Eq.~(\ref{nuimuifree}) for free-particle systems should be
modified to include interactions. In fact for the quark flavor $i$ we have
\begin{equation}
\mu_i
= \left.
   \frac{\mathrm{d} E}{\mathrm{d} n_i}
  \right|_{\{n_{k\neq i}\}}
=\frac{\partial E_i}{\partial \nu_i}  \frac{\mathrm{d}\nu_i}{\mathrm{d} n_i}
 +\sum_j \frac{\partial E}{\partial m_j}\frac{\partial m_j}{\partial n_i}.
\label{quasi}
\end{equation}
Since the quark masses are density dependent, the derivatives generate an
additional term with respect to the free Fermi gas model. We get
\begin{equation}
\mu_i =
  \frac{n_i}{|n_i|}\sqrt{\nu_i^2+m_i^2}
  +\sum_j |n_j|\frac{\partial m_j}{\partial n_i}
   f\!\left(\frac{\nu_j}{m_j}\right),
 \label{mui}
\end{equation}
where
\begin{equation}
f(x) \equiv \frac{3}{2x^3} \left[
 x\sqrt{1+x^2}-\ln\left(x+\sqrt{1+x^2}\right)
\right].
\end{equation}
 The pressure is then given by
\begin{eqnarray}
P&=& 
    -E + \sum_i \mu_i n_i
\nonumber \\
&=&
  -\Omega_0
  +\sum_{ij} n_i|n_j|\frac{\partial m_j}{\partial n_i}
   f\left(\frac{\nu_j}{m_j}\right),
 \label{pressure}
\end{eqnarray}
with $\Omega_0$\ being the free-particle contribution:
\begin{eqnarray}
\Omega_0
&=&
-\sum_i\frac{g_i}{48\pi^2}
\left[
 \nu_i\sqrt{\nu_i^2+m_i^2}\left(2\nu_i^2-3m_i^2\right)
\right.
\nonumber\\
&& \phantom{-\sum_i\frac{g_i}{48\pi^2}[}
 \left.
 +3m_i^4\,\mbox{arcsinh}\left(\frac{\nu_i}{m_i}\right)
\right].
\end{eqnarray}

Due to the additional term in the chemical potential,
the pressure also has an extra term. The inclusion of such a
term guarantees  that the Hughenoltz-Van Hove theorem is
fulfilled in the calculations.

In the quasi-particle model \cite{PeshierPRC61}, one also has an extra term.
Because the quark masses there depend on chemical potentials, and
the extra term is not used in the relation between the Fermi momenta
and chemical potentials, an effective bag constant has to be added
to the energy expression to consider confinement \cite{SchertlerNPA616}.

In CDDM quark model, however, we do not need a bag constant any more.
Quark confinement is achieved automatically by the density dependence
of quark masses, or by the strong interaction between quarks. In fact,
the exponent $z=1/3$ in Eq.\ (\ref{mqT0}) is derived from the linear
confinement interaction  \cite{PenggxPRC61}.

\begin{figure}[htb]
\centering
\includegraphics[width=8.2cm]{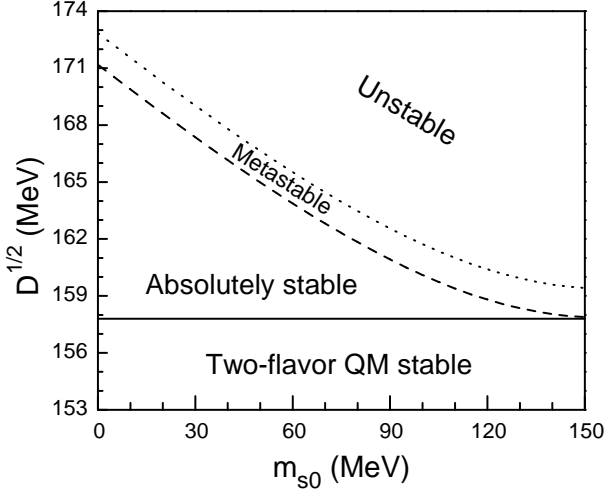}
\caption{ Confinement constant range determined by stability
arguments. Only in the region bounded by the full and dashed lines,
SQM is absolutely stable.
         }
 \label{Dms0}
\end{figure}

In the present model, the parameters are: the electron mass
$m_e=0.511$ MeV, the quark current masses $m_{u0}$, $m_{d0}$, $m_{s0}$,
and the confinement parameter $D$.
Although the light-quark masses are not without controversy and
remain under active investigations, they are anyway very small,
and so we simply take $m_{u0}=m_{d0}=0.$
The current mass of strange quarks is $95\pm 25$ MeV according
to the latest version of the Particle Data Group \cite{PDG}.

Conventionally, the stability of strange quark matter
(SQM) is judged by the minimum energy per baryon
\cite{WittenPRD,Jaffe,ChakrabartyPLB229,BenvenutoPRD51}.
If it is less than 930 MeV (the mass of $^{56}$Fe divided by 56),
then SQM is absolutely stable. If it is bigger than 930 MeV
but less than 939 MeV (the mass of nucleons), then it is metastable.
Otherwise, if it is larger than 939 MeV, SQM is un-stable.
In case of two-flavor quark matter,
it should be no less than 930 MeV, in order not to contradict
standard nuclear physics.
This is the Witten-Bodmer hypothesis \cite{WittenPRD,Jaffe}.

in Fig.~\ref{Dms0}, we show the different regimes
in the $\sqrt{D}$-$m_{s0}$ plane.
The area below the full line is forbidden where the energy
per baryon of two-flavor quark matter is less than
930 MeV. Above the dotted line, the energy per baryon of
SQM is more than 939 MeV, and thus SQM is unstable.
The area bounded by the dotted and dashed lines is
the metastable region where the energy per baryon is
between 930 MeV and 939 MeV. Only when the $(D^{1/2},m_{s0})$ pair
is in the range between the full and dashed lines, SQM
can be absolutely stable, i.e., its energy per baryon
is less than 930 MeV. Therefore, the range of $D$ values is
very narrow for a chosen $m_{s0}$ value, if the Witten-Bodmer hypothesis
is correct. If we take the modest value $m_{s0}=100$ MeV, for example,
then $D^{1/2}$ is in the range of 158--160 MeV.
The lower bound 158 MeV is obtained by taking $m_{u0}=m_{d0}=0$.
If $m_{u0}$ and $m_{d0}$ are given a small finite value,
the lower bound can then be a little bit smaller, e.g. 156 MeV \cite{PenggxPRC62}.

Unfortunately we presently do not have a definite conclusion
on the stability of SQM, so we treat $D$ as a free parameter.
However, the first condition, i.e.,
$D$ greater than about $(158\ \mbox{MeV})^2$, should always be satisfied.
On the other hand, we can connect $D$ to the
pion mass $m_{\pi}$, pion decay constant $f_{\pi}$, pion-nucleon sigma term
$\sigma_{\mathrm{N}}$, string tension $\sigma_0$,
and the vacuum chiral condensate $\langle\bar{q}q\rangle_0$ by
\cite{WenxjPRC72}
\begin{equation}
D=\frac{3(2/\pi)^{1/3}\sigma_0m_{\pi}^2f_{\pi}^2}
       {-\sigma_{\mathrm{N}}\sum_q\langle\bar{q}q\rangle_0}.
\end{equation}
 From the known range of the vacuum condensate, we can have
an upper bound $(270\ \mbox{MeV})^2$.
Therefore, $D^{1/2}$ should not be out of the range (156, 270) MeV.

\subsection{Properties of strange quark matter}
\label{propSQM}

As usually done, we consider SQM as a mixture of $u$, $d$, $s$
quarks, and electrons. The relevant chemical potentials
$\mu_u$, $\mu_d$, $\mu_s$, and $\mu_e$ satisfy the
weak-equilibrium condition
\begin{eqnarray}
& \mu_u+\mu_e=\mu_d, &
\label{weak1} \\
& \mu_d=\mu_s. &
\label{weak2}
\end{eqnarray}

Because all particle masses do not depend on the density of electrons,
i.e., $\partial m_j/\partial n_e=0$, Eq.~(\ref{mui}) gives
\begin{equation}
\mu_i
=\sqrt{(\pi^2n_i)^{2/3}+m_i^2}-\mu_{\mathrm{I}}
\label{muivsni}
\end{equation}
with
\begin{equation}
\mu_{\mathrm{I}}
=-\frac{1}{3}\frac{\partial m_{\mathrm{I}}}{\partial n_{\mathrm{b}}}
  \sum_{j=u,d,s} n_jf\left(\frac{\nu_j}{m_j}\right)
\label{muIvsni}
\end{equation}
for $i=u,d,s$ quarks, and
\begin{equation} \label{muevsne}
\mu_e=\sqrt{\left(3\pi^2n_e\right)^{2/3}+m_e^2}
\end{equation}
for electrons.

In Eq.~(\ref{muIvsni}), $m_{\mathrm{I}}$ is the
second term on the right hand side of Eq.~(\ref{mqT0}),
so we have
$
\partial m_{\mathrm{I}}/\partial n
=-zD/n^{z+1}
=-zm_{\mathrm{I}}/n.
$
The pressure is then obtained from Eq.\ (\ref{pressure}) as
\begin{equation} \label{Psqm}
P=-\Omega_0
  +n_{\mathrm{b}}\frac{\mathrm{d}m_{\mathrm{I}}}{\mathrm{d}n_{\mathrm{b}}}
   \sum_{j=u,d,s} n_j\ f\!\left(\frac{\nu_j}{m_j}\right).
\end{equation}

Substituting these expressions into Eqs.~(\ref{weak1}) and (\ref{weak2}),
we have
\begin{eqnarray}
&& \sqrt{\left(\pi^2n_u\right)^{2/3}+m_u^2}\
 +\sqrt{\left(3\pi^2n_e\right)^{2/3}+m_e^2}
\nonumber\\
&&
=\sqrt{\left(\pi^2n_d\right)^{2/3}+m_d^2}.
\label{qmeq1}
\end{eqnarray}
and
\begin{equation} \label{qmeq2}
 \left(\pi^2n_d\right)^{2/3}+m_d^2
=\left(\pi^2n_s\right)^{2/3}+m_s^2.
\end{equation}
We also have the baryon number density
\begin{equation} \label{qmeq3}
n=\frac{1}{3}(n_u+n_d+n_s)
\end{equation}
and the charge density
\begin{equation} \label{qmeq4}
Q_{\mathrm{q}}=\frac{2}{3}n_u-\frac{1}{3}n_d-\frac{1}{3}n_s-n_e.
\end{equation}
The charge-neutrality condition requires $Q_{\mathrm{q}}=0$.

For a given total baryon number density $n$, we can
obtain the respective $n_u$, $n_d$,  $n_s$, and $n_e$ by solving the
four equations (\ref{qmeq1}), (\ref{qmeq2}), (\ref{qmeq3}),
(\ref{qmeq4}). The chemical potentials $\mu_u$, $\mu_d$,  $\mu_s$,
and $\mu_e$ can then be calculated by Eqs.\ (\ref{muivsni}) and
(\ref{muevsne}).
Therefore, the energy density of the quark matter is a function of
the baryon number density $n$ and the charge density $Q_{\mathrm{q}}$,
i.e., $E_{\mathrm{q}}=E_{\mathrm{q}}(n,Q_{\mathrm{q}})$ or
\begin{equation}  \label{dEqbaryonchem}
\mbox{d}E_{\mathrm{q}}
=\frac{\partial E_{\mathrm{q}}}{\partial n} \mbox{d}n
 +\frac{\partial E_{\mathrm{q}}}{\partial Q_{\mathrm{q}}} \mbox{d}Q_{\mathrm{q}},
\end{equation}
where the two partial derivatives,
$\partial E_{\mathrm{q}}/\partial n$
and
$\partial E_{\mathrm{q}}/\partial Q_{\mathrm{q}}$
are called the baryon chemical potential and charge chemical potential,
respectively. It can be easily shown that they are connected to the
quark chemical potentials by
\begin{equation}  \label{baryonchemmuud}
\frac{\partial E_{\mathrm{q}}}{\partial n}
=\mu_u+2\mu_d, \ \
\frac{\partial E_{\mathrm{q}}}{\partial Q_{\mathrm{q}}}
=\mu_u-\mu_d.
\end{equation}

In fact, according to the fundamental differentiation equality
of thermodynamics, we have
\begin{equation} \label{dEqthermo}
\mbox{d}E_{\mathrm{q}}
= \mu_u\mbox{d}n_u
 +\mu_d\mbox{d}n_d
 +\mu_s\mbox{d}n_s
 +\mu_e\mbox{d}n_e.
\end{equation}
On the other hand, we have $\mu_e=\mu_d-\mu_u$, $\mu_s=\mu_d$,
$n_s=3n-n_u-n_d$, and $n_e=n_u-n-Q_{\mathrm{q}}$ from
Eqs.~(\ref{weak1}), (\ref{weak2}), (\ref{qmeq3}), and (\ref{qmeq4}).
Substituting these four equalities into Eq.~(\ref{dEqthermo}) leads
to
$\mbox{d}E=(\mu_u+2\mu_d)\mbox{d}n+(\mu_u-\mu_d)\mbox{d}Q_{\mathrm{q}}$.
Comparison of this with Eq.~(\ref{dEqbaryonchem}) immediately gives
Eq.~(\ref{baryonchemmuud}).

\begin{figure}[htb]
\centering
\includegraphics[width=8.2cm]{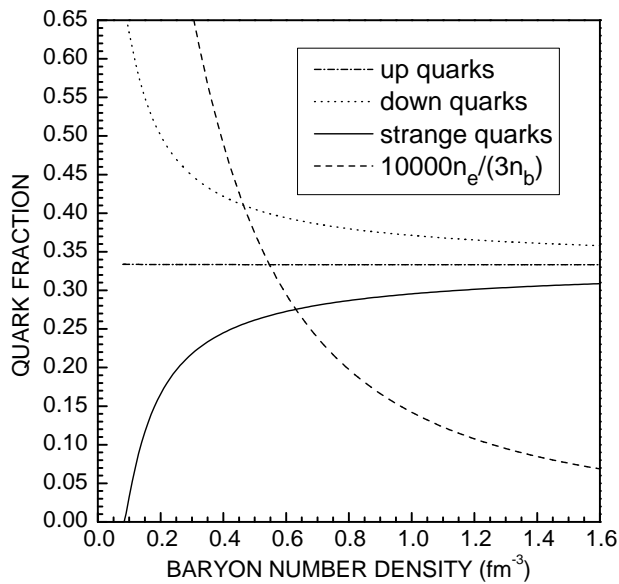}
\caption{
 Quark fraction vs baryon number density for $D^{1/2}=160$ MeV
 and $m_{s0}=80$ MeV.
         }
 \label{nVSnb}
\end{figure}

In Fig.~\ref{nVSnb}, the quark fractions, i.e., $n_u/(3n)$,
$n_d/(3n)$, $n_s/(3n)$, and the $10^4$ times the electron number
divided by the total quark number, $10000n_e/(3n)$, have been shown
versus the baryon number density for $D^{1/2}=160$ MeV and
$m_{s0}=80$ MeV. It is seen that the fraction of up quarks is nearly
always one third. The fraction of down quarks increases rapidly with
decreasing densities, while the fraction of strange quarks
approaches to zero when the density decreases to a certain lower
density.

\begin{figure}[htb]
\centering
\includegraphics[width=8.2cm]{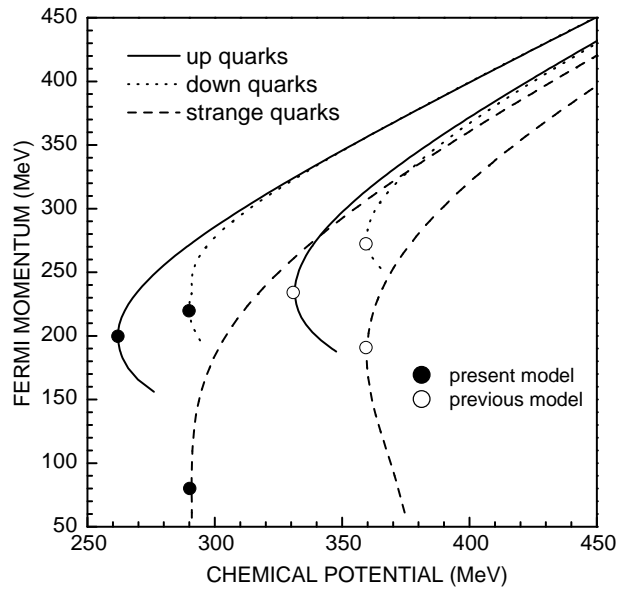}
\caption{
 Comparison between the Fermi momenta
 and chemical potentials.
 Parameters are the same as for Fig.\ \ref{nVSnb}.
         }
 \label{nui-mu}
\end{figure}

In order to compare the relation between the Fermi momentum and chemical
potential, we plot, in Fig.~\ref{nui-mu}, the Fermi momentum of up (solid line),
down (dotted line), and strange (dashed line) quarks, respectively,
as a function of the corresponding quark chemical potential, in both
the present model (lines with a solid circle) and the previous model
(lines with an open circle). It is very obvious that the difference is
very large, especially at comparatively lower densities. In both models,
the Fermi momentum of up or down quarks is higher than that of strange
quarks due to the fact that strange quarks are heavier than up or down
quarks. For the same chemical potential, however, the Fermi momentum in
the present model is generally bigger than that in the previous model,
due to the quark mass-density-dependence which reflects the strong
interaction between quarks.

\begin{figure}[htb]
\centering
\includegraphics[width=8.2cm]{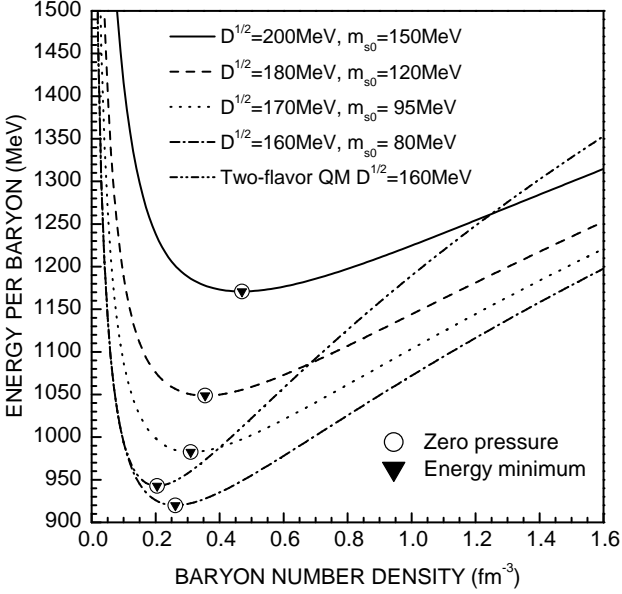}
\caption{
 Energy per baryon of quark matter in the present model.
 The parameter pair $(D^{1/2},m_{s0})$ in MeV for
 the solid, dashed, dotted, and dash-dotted cures are
 (200,150), (180,120), (170,95), and (160,80), respectively.
 The dash-dot-dot line is for the two-flavor quark matter
 at $D^{1/2}=160$ MeV. It is very obviously shown that
 the energy minimum, marked with a full triangle on each line,
 is located exactly at the same point of the zero pressure
 indicated by an open circle.
         }
 \label{Enb}
\end{figure}

Figure \ref{Enb} shows the energy per baryon of quark matter for
different parameter sets in the present model. Each line has a
minimum, corresponding to the lowest energy state (marked with a
solid triangle). One can see that the pressure at this minimum is
exactly zero. So this special point is marked with an open circle as
well. At the same time, we also display the energy per baryon for
the two-flavor quark matter by a dash-dot-dot line. We see that the
two-flavor quark matter is less stable than SQM. Even for a smaller
$D$ value, e.g., $(160\ \mbox{MeV})^2$, its energy will finally
exceed that of SQM for a bigger $D$ with increasing densities.

\begin{figure}[htb]
\centering
\includegraphics[width=8.2cm]{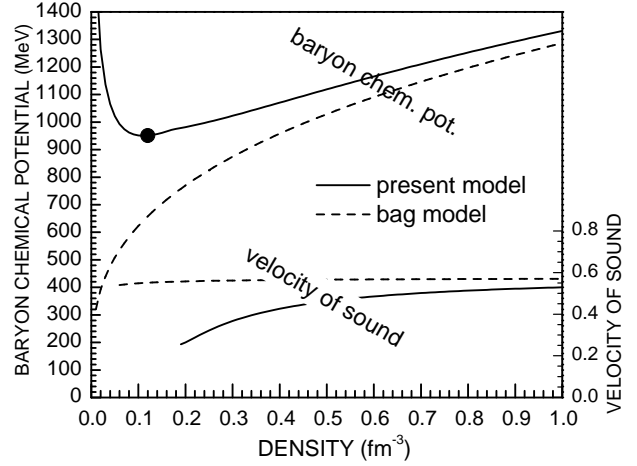}
\caption{The baryon chemical potential in CDDM ($\sqrt{D}=180$ MeV,
$m_{s0}=120$ MeV) and in the bag model ($B^{1/4}=180$ MeV). The
former has a minimum at a lower density depending on the value of
$D$ and $m_{s0}$, while the latter is always a monotonic function of
density. The velocity of sound in both models has also been given on
the right axis.
         }
 \label{comp}
\end{figure}

In principle, the CDDM quark model contains more physics than the simple
bag model. To demonstrate this we plot, in Fig.~\ref{comp}, the
baryon chemical potential in CDDM with $D^{1/2}=180$ MeV and
$m_{s0}=120$ MeV, and in the bag model with $B^{1/4}=180$ MeV. In
CDDM, the baryon chemical potential decreases with decreasing
density to a certain value depending on $D$ and $m_{s0}$, then it
increases very rapidly, i.e., it saturates at a definite density
marked with a bullet. When the density is lower than the bullet, the
derivative $\mbox{d}^2E/\mbox{d}n^2$ becomes negative, and so quark
matter is unstable against phase separation and falls apart at lower
densities. In the bag model, however, the baryon chemical potential
is always a monotonic function of density, which means that quark
matter does not fall apart at any lower densities. The velocity of
sound has also been plotted on the right axis. We observe that it
is, in the bag model, nearly the same as that of a non-interacting
Fermi gas. But in CDDM, it decreases to zero with decreasing densities.
At high densities, it becomes asymptotically identical to the
ultrarelativistic limit, as expected.

\subsection{Quark matter at finite temperature}
\label{finiteT}

Since the single particle energies depend on density and temperature
via the quark masses, the thermal properties should be founded on
the canonical ensemble, but, as is well known, the partition
function is not easy to calculate. Therefore a different statistical
procedure is usually adopted, which is based on the quasiparticle
assumption. According to that the energy density is written as
\begin{eqnarray}
E &=& \sum_i{g_i} \sum_{\bf p} \sqrt{p^2+m_i^2}\,\,\mbox{f}_i(p,T),
\label{ass1}
\end{eqnarray}
where the Fermi distribution function is
\begin{equation}
\mbox{f}_i(p,T)
=\frac{1}{1+e^{[\epsilon_i(p,T)-\mu_i]/T}}.
\label{ass3}
\end{equation}
If antiparticles are included, the sum must be extended to
antiparticles for which $\mu_i$ must be replaced by $-\mu_i$.
From the Landau definition of the single-particle energy extended
to finite temperature, we have
\begin{eqnarray}
\epsilon_i(p)
 &=& \frac{\delta E}{\delta \mbox{f}_i(p,T)}
 \nonumber \\
 &=&
    \sqrt{p^2+m_i^2}
     +\sum_j {g _j}\frac{m_j \mbox{f}_j(p,T)}{\sqrt{p^2+m_j^2}}
      \frac{\partial m_j}{\partial n_i}
 \nonumber \\
&\equiv& \varepsilon_i(p)-\mu_{\mathrm{I}},
\end{eqnarray}
where $\varepsilon_i(p)\equiv\sqrt{p^2+m_i^2}$\ is the dispersion
relation of free particles.
The extra term $\mu_{\mathrm{I}}$ can be added to the chemical
potential, so defining
\begin{equation}
\mu_i^*\equiv\mu_i+\mu_{\mathrm{I}}.
\end{equation}
Accordingly, the net density of the particle type $i$ is
$
n_i=g_i\sum_{\bf p} \left[\mbox{f}_i(p,T)
    -\mbox{f}_{\bar{i}}(p,T)\right],$\
or, explicitly, we have
\begin{equation}
n_i
= g_i\int_0^{\infty}
  \left\{\frac{1}{1+e^{[\varepsilon_i(p)-\mu^*_i]/T}}\right.
  \left.-\frac{1}{1+e^{[\varepsilon_i(p)+\mu^*_i]/T}}\right\}
  \frac{p^2\mbox{d}p}{2\pi^2}.
\end{equation}
Inverting this equation, one determines $\mu^*_i$ as a function
of $n_i$ so that the free energy density
\begin{equation}
F
=\sum_iF_i(T,\mu_i^*,m_i)
=\sum_i
 \left[F_i^++F_i^-\right]
\label{FTtot}
\end{equation}
with
\begin{eqnarray}
F_i^{\pm}
&=&
 g_i
 \int_0^{\infty}
 \bigg\{
 -T\ln\left[1+e^{-(\sqrt{p^2+m_i^2}\mp\mu_i^*)/T}
    \right]
\nonumber\\
&& \phantom{g_i\int_0^{\infty}\bigg\{}
 \pm\frac{\mu_i^*}{1+e^{(\sqrt{p^2+m_i^2}\mp\mu_i^*)/T}}
 \bigg\} \frac{p^2\mbox{d}p}{2\pi^2}
 \label{FiTmum}
\end{eqnarray}
will be a function of respective particle densities, instead of
chemical potentials. One can then determine the real chemical
potentials and pressure, according to the well-known relations
\begin{eqnarray}
\mu_i = \frac{\partial F}{\partial n_i}, \ \
P     =-F+\sum_i\mu_in_i.
\end{eqnarray}
These quantities will completely describe the thermal equilibrium
of pure quark phase and the transition to the quark-hadron mixed phase.
A more detailed analysis, where the thermodynamic formulism is
developed to much more extent, is reported
in Appendix~\ref{appeA}. In Ref.~\cite{WenxjPRC72},
the $\mu_i$ should be implicitly understood as $\mu_i^*$.

\begin{figure}[htb]
\centering
\includegraphics[width=8.2cm]{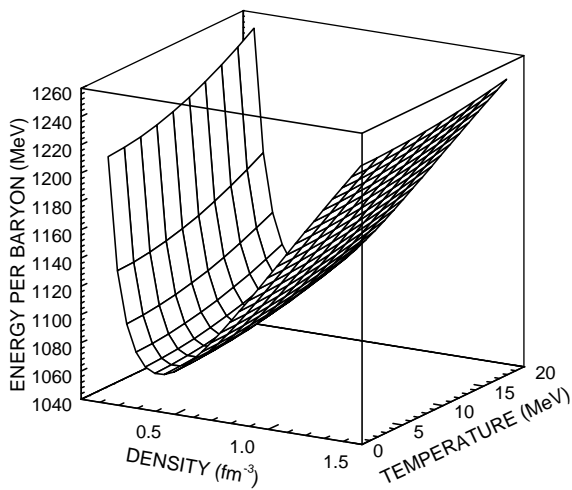}
\caption{
 Density and temperature dependence of the energy per
baryon for the parameters $\sqrt{D}=180$ MeV and $m_{s0}=120$ MeV.
         }
 \label{EnbT}
\end{figure}

In Fig.~\ref{EnbT}, we give the energy per baryon
of SQM as a function of both density and temperature.
The parameters used for this three-dimensional plot
are $\sqrt{D}=180$ MeV and $m_{s0}=120$ MeV.

\section{Nuclear matter in Brueckner theory with three-body forces}
\label{BruecknerEOS}

The Brueckner-Bethe-Goldstone (BBG) theory is among the most advanced microscopic theories of
nuclear matter. In the recent years it was recognized that the three-body forces, which are
expected to have a dominant role at high nuclear density, also affect the saturation point and,
in fact, after including the three-body forces in the Brueckner theory, the empirical
saturation properties are reproduced quite well~\cite{Zuowei02prc,zhli}. In the neutron star
interior, where high baryonic density values are reached, processes like the excitation of
nucleon-antinucleon pairs ($Z$-diagrams) and nucleonic resonances (together with the production
of other hadrons) sizeably influence the two-body nuclear interaction.
The former process involves the virtual excitation of negative
energy states, which is absent from the standard Brueckner
theory, and thus it represents a pure relativistic effect.
From the comparison with the Dirac-Brueckner theory it turns
out that it is by far the most important relativistic effect \cite{zhli}.
This process together with nucleon resonances can be incorporated
in the interaction as medium virtual excitations in TBFs.
One could guess that many body (more that three) forces are also
important at high density as large as $\rho\approx 1\
\mbox{fm}^{-3}$, but it is hard to imagine that pure baryon matter
can exist at so high density. At lower density two and three-body
forces are dominant since the hole line expansion is, roughly
speaking, an expansion in density powers.

The global effect of TBFs at
high density is strongly repulsive, leading to a remarkable increase of the maximum mass in the
study of the neutron star structure. But also a correct estimate of the saturation point is
important, since, as we will see below, in strongly asymmetric nuclear matter the threshold for
the transition to mixed nucleon-quark phase can appear close to the saturation density.
Therefore the corresponding EoS could be used as input for
transport-model simulations of heavy ion collisions, where strongly
isospin-asymmetric systems are formed in central events.

\subsection{BBG equations}

The Brueckner theory extended to TBF is described elsewhere~\cite{grange:1989,Zuowei02prc}.
Here we simply give a brief review of the Brueckner-Hartree-Fock (BHF) approximation at finite
temperature $T$~\cite{cugnon,baldo,Zuoweit}. The starting point is the reaction $G$-matrix, which
satisfies the BBG equation,
\begin{equation}
G(\omega,T)=\upsilon_{\mathrm{NN}} +\upsilon_{\mathrm{NN}} \sum_{k_{1}k_{2}}\frac{
|k_{1}k_{2}\rangle Q_{k_{1},k_{2}}(T)\langle k_{1}k_{2}|}{\omega -\epsilon_{k_{1}}(T)-\epsilon
_{k_{2}}(T)}G(\omega,T),
\end{equation}
where $k_i\equiv(\vec k_i,\sigma_i,\tau_i)$, denotes the single particle momentum, the
$z$-component of spin and isospin, respectively, and $\omega$ is the starting energy. The
$G$-matrix, the Pauli operator $Q$ and the single particle energies
$\epsilon_k(T)=k^2/2m+U_k(T)$ depend on the neutron and proton densities and temperature. The
interaction $\upsilon_{\mathrm{NN}}$ given by
\begin{equation}
\upsilon_{\mathrm{NN}} = V^{\mathrm{bare}}_2 + V^{\mathrm{eff}}_3,
\end{equation}
where  $V^{\mathrm{bare}}_2$ is the bare two body force (2BF) and
$V^{\mathrm{eff}}_3$ is an effective 2BF derived by the average of
the bare TBF on the third particle as follows
\begin{equation}
\begin{array}{lll}
 \langle\vec r_1 \vec r_2|V^{\mathrm{eff}}_3(T)|\vec r_1^{\ \prime} \vec r_2^{\ \prime} \rangle =
\\[2mm]\displaystyle
 \frac{1}{4}
{\rm Tr}\sum_n \int {\rm d} {\vec r_3} {\rm d} {\vec r_3^{\ \prime}}\phi^*_n(\vec r_3^{\
\prime}) [1-\eta(r_{13}',T )]
[1-\eta(r_{23}',T)] \\[6mm]
\times \displaystyle W_3(\vec r_1^{\ \prime}\vec r_2^{\ \prime} \vec r_3^{\ \prime}|\vec r_1
\vec r_2 \vec r_3) \phi_n(r_3) [1-\eta(r_{13},T)][1-\eta(r_{23},T)].
\end{array}
\label{eq:TBF}
\end{equation}

Since the defect function $\eta(r,T)$ is directly determined by the
solution of the BBG equation \cite{grange:1989},
$V^{\mathrm{eff}}_3$ must be calculated self-consistently with the
$G$ matrix and the s.p.\ potential $U_k$ on the basis of BBG
equation. It is clear from Eq.~(\ref{eq:TBF}) that the effective
force rising from the TBF in nuclear medium is density and
temperature dependent through the defect function. A detailed
description and justification of the method can be found in
Ref.~\cite{grange:1989},
including a discussion on the averaging procedure. The validity
of such a procedure has been numerically tested in the comparison
between the BHF EoS plus Z-diagrams with $\sigma$ meson exchange
and the Dirac-BHF EoS, which are expected to be equal. The
calculation \cite{zhli} gives an impressive agreement between the
two EoS's, although the TBF due to the $Z$-diagrams is averaged
according to Eq.~(\ref{eq:TBF}).

 For $V^{\mathrm{bare}}_2$ we adopt the Argonne
$V_{18}$ two-body interaction~\cite{wiringa:1995}. The TBF is constructed from the
meson-exchange current approach ~\cite{grange:1989} and contains virtual particle ($\Delta$ and
$N^*(1440)$) excitations and, in addition, relativistic effects induced by the excitations of
particle-antiparticle pairs. This description of the interaction is not completely consistent
since, in principle, the two and three body forces should be derived from the same meson
parameters, but a recent calculation \cite{zhli} replacing Argonne potential with Bonn
potential \cite{bonn} and the TBF built up with the Bonn meson parameters
substantially leads to the same results.

\begin{figure}[htb]
\centering
\includegraphics[width=8.2cm]{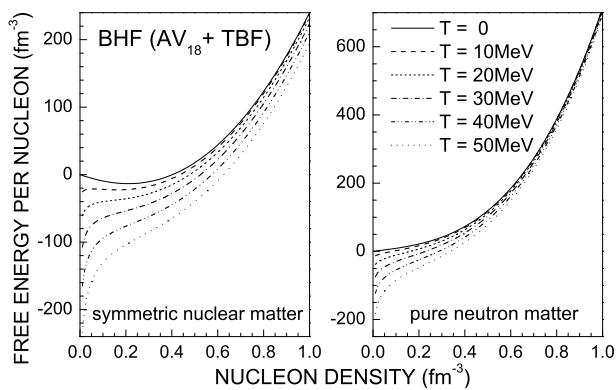}
\caption{Isotherms of symmetric nuclear matter (left side) and pure
neutron matter (right side) as a function of the nucleon density at
different temperature.} \label{eoslast}
\end{figure}

\subsection{Thermodynamics}

Let us start with symmetric nuclear matter. In the BHF
approximation, the thermodynamic potential can be written
\begin{eqnarray}
\Omega_{\mathrm{N}}
 &=& \Omega_{\mathrm{N}}^{0} + W_L, \\
\Omega_{\mathrm{N}}^{0}
 &=& -T\sum _k \ln\left[1+e^{-(\epsilon_k-\mu)/T}\right], \\
W_L&=&-\frac{1}{2}\sum_k f_k(T) U_k(T),
\end{eqnarray}
where $\Omega_{\mathrm{N}}^{0}$ is the thermodynamic potential for a
system of independent particle with single particle spectrum
$\epsilon_k=\frac{\hbar^2 k^2}{2m}+ U_k(T)$, and the $W_L$ the sum
of all linked cluster diagrams to the lowest order in the hole line
expansion\cite{cugnon}. $U_k(T)$ is the selfconsistent mean field,
\begin{equation}
U_k(T)\,=\, \sum_{k'} \langle kk'|G|kk'\rangle_{\mathrm{A}}
f_{k'}(T).
\end{equation}
The finite temperature BHF approximation suffers from the same
difficulty as any strongly interacting Fermi system. The difficulty
is the same as in the CDDM quark model, since in both cases the s.p.
spectrum is density and temperature dependent. Whereas at $T=0$ the
density (or Fermi momentum) can be fixed, at $T>0$ the role of
density is taken by chemical potential $\mu$ (grand canonical
ensemble), and one is forced to fix the chemical potential at each
iteration due to the presence of the Fermi distribution
$f_k(T)=\{1+\exp([\epsilon_k(T)-\mu]/T)\}^{-1}$. Since this
procedure does not converge \cite{cugnon}, one should fix the
density and invert the equation relating density and chemical
potential,
\begin{equation}
\rho =\frac{1}{V}\sum_k f_k(T)
 -\left(
   \frac{\partial W_L}{\partial \mu}
        \right)_T,
\end{equation}
which is not a viable task. The usual approximation is to drop out
the derivative in the previous equation, which corresponds to the
quasiparticle approximation above discussed within the CDDM quark
model. So doing, the resulting $\tilde\mu$ looses its meaning of
chemical potential. In this approximations the energy and entropy
densities are given by:
\begin{eqnarray}
E_N &=& \frac{1}{V}\sum_k f_k(T)
       \left[\frac{\hbar^2 k^2}{2m}+ U_k(T)\right], \\
S_N &=& -\frac{1}{V}\sum_k\Big\{f_k(T)\ln f_k(T)
\nonumber\\
& & \phantom{-\sum_k\Big\{}
  +\big[1-f_k(T)\big] \ln\big[1-f_k(T)\big]\Big\}.
\end{eqnarray}
After one calculates $\tilde\mu$ in terms of $\rho$, the
thermodynamics is developed from the free energy density
$F_{\mathrm{N}}(\rho,T)= E_{\mathrm{N}}(\rho,T)-TS_N(\rho,T)$. The
free energy per particle, calculated from BHF approximation, is
depicted in Fig.~\ref{eoslast}. Due to the difficulty of extending
the BHF code to very high temperature an extrapolation from the real
numerical results to high $T$ has been performed adopting the so
called frozen approximation based on $T$-independent single particle
spectrum, i.e.\ the latter is frozen at $T=0$. This turns out to be
a good approximation up to $10\sim 20$ MeV \cite{balfer}.

The relevant thermodynamical quantities, i.e.\ chemical potentials
and pressure, are derived from free energy as follows
\begin{equation}
\mu = \left.
   \frac{\partial F_N}{\partial \rho}
  \right|_T,  \ \ \
P = \rho^2 \frac{\mbox{d}(F_N/\rho)}{\mbox{d} \rho}.
\end{equation}

Let us consider asymmetric nuclear matter with baryon density
$\rho=\rho_{\mathrm{n}}+\rho_{\mathrm{p}}$ and asymmetry parameter
$\beta=(\rho_{\mathrm{n}}-\rho_{\mathrm{p}})/\rho$, where
$\rho_{\mathrm{n}}$ ($\rho_{\mathrm{p}}$) is the neutron (proton)
density. The baryon chemical potentials can be expressed as
\begin{eqnarray}
\mu_{\mathrm{n}}
= \left(\frac{\partial
F_{\mathrm{N}}}{\partial\rho_n}\right)_{T,\rho_{\mathrm{p}}}, \ \ \
\mu_{\mathrm{p}} = \left(\frac{\partial
F_{\mathrm{N}}}{\partial\rho_{\mathrm{p}}}\right)_{T,\rho_{\mathrm{n}}},
\end{eqnarray}
where $F_{\mathrm{N}}$ is the free energy density.

Assuming the parabolic law for the latter, we get the simple expression
for the chemical momentum isotopic shift
\begin{equation} \label{esym}
\mu_{\mathrm{n}}-\mu_{\mathrm{p}}= 4\beta F_{\mathrm{sym}}(\rho,T),
\end{equation}
where $F_{\mathrm{sym}}$ is the symmetry free energy density.
The parabolic law is well satisfied at low density, but at high
density additional terms of the $\beta$-expansion must be considered.

In neutron star inner core nuclear matter is supposed to be in
$\beta$-equilibrium under the condition of charge neutrality.
Assuming that only electrons are present (the muon contribution
is negligible), the two preceding conditions require
\begin{eqnarray}
\mu_e         &=& \mu_{\mathrm{n}}-\mu_{\mathrm{p}}, \\
Q_{\mathrm{N}}&=& \rho_{\mathrm{p}}- \rho_e.
\end{eqnarray}
$Q_{\mathrm{N}}$ is the net charge density of nuclear matter.
It is zero for pure neutral nuclear matter.
For a given set of $(\rho,Q_{\mathrm{N}})$, we can solve
the chemical potentials $\mu_{\mathrm{n}}$, $\mu_{\mathrm{p}}$,
and $\mu_e$ from the above equations. Then all other quantities
can be obtained for a fixed temperature. In other words,
all thermodynamic quantities can be regarded as a function
of the nucleon density $\rho$, charge density $Q_{\mathrm{N}}$,
and temperature $T$. At zero temperature, for example,
the energy density can be regarded as a function of $\rho$ and
$Q_{\mathrm{N}}$, i.e.,
$E_{\mathrm{N}}=E_{\mathrm{N}}(\rho,Q_{\mathrm{N}})$.
With a similar approach as in the preceding section to obtain
Eq.~(\ref{baryonchemmuud}), we can easily show that
the baryon chemical and charge chemical potentials of
nuclear matter can be expressed as
\begin{equation} \label{baryonchemNM}
\frac{\partial E_{\mathrm{N}}}{\partial \rho_n} =\mu_{\mathrm{n}},\ \ \frac{\partial
E_{\mathrm{N}}}{\partial Q_{\mathrm{N}}} =\mu_{\mathrm{p}}-\mu_{\mathrm{n}}.
\end{equation}

\begin{figure}[htb]
\centering
\includegraphics[width=8.2cm]{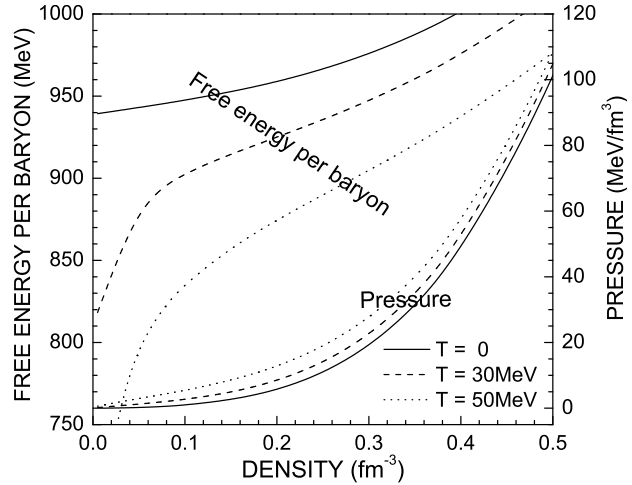}
\caption{
 The free energy per baryon (left y axis) and pressure (right y axis)
 of nuclear matter in $\beta$-equilibrium. Three values of the temperature
 are considered.}
 \label{beos}
\end{figure}

The system turns out to be in a strongly isospin asymmetric state.
The isotherms of free energy and
pressure of nuclear matter in $\beta$-equilibrium are shown
in Fig.~\ref{beos}.

\section{Phase diagram structure at zero and finite temperature}
\label{phase}

Let us study the nuclear matter, consisting of nucleons and electrons,
in equilibrium with a gas of $u$, $d$, $s$ quarks and electrons.
According to Glendenning~\cite{glen1,glen2}, we assume the total
charge conservation, in addition to total baryon and energy conservation.
Now we first consider the case of zero temperature, and then extend to
finite temperature.

The conservation laws can be imposed by introducing the quark
fraction $\chi$ defined as
\begin{equation}
\chi\equiv V_{\mathrm{q}}/V.
\end{equation}
where $V$ is the total volume, $V_{\mathrm{q}}$ is the volume occupied by quarks.
Then the total baryon density is
\begin{equation} \label{rhotot}
\rho_{\mathrm{t}}
 = (1-\chi) \rho +\chi n,
\end{equation}
the total electric charge is
\begin{equation} \label{Qtot}
Q_{\mathrm{t}}
 = (1-\chi) Q_{\mathrm{N}} +\chi Q_{\mathrm{q}},
\end{equation}
and the total energy density is
\begin{equation} \label{Etot}
E_{\mathrm{t}} = (1-\chi) E_{\mathrm{N}} +\chi E_{\mathrm{q}},
\end{equation}
where $\rho$, $Q_{\mathrm{N}}$, and $E_{\mathrm{N}}$ are, respectively, the baryon number
density, electric charge density, and energy density of nuclear matter, while $n$,
$Q_{\mathrm{q}}$, and $E_{\mathrm{q}}$ are the corresponding quantities of quark matter.
$E_{\mathrm{N}}$ is a function of $\rho$ and $Q_{\mathrm{N}}$, $E_{\mathrm{q}}$ is a function
of $n$ and $Q_{\mathrm{q}}$, i.e., $E_{\mathrm{N}}=E_{\mathrm{N}}(\rho,Q_{\mathrm{N}})$,
$E_{\mathrm{q}}=E_{\mathrm{q}}(n,Q_{\mathrm{q}})$.
Differentiating Eq.~(\ref{Etot}), one obtains
\begin{eqnarray}
\mbox{d}E_{\mathrm{t}}
&=&
 (1-\chi)
 \left(
  \frac{\partial E_{\mathrm{N}}}
       {\partial \rho}
  \mbox{d}\rho
  +\frac{\partial E_{\mathrm{N}}}
   {\partial Q_{\mathrm{N}}}
   \mbox{d}Q_{\mathrm{N}}
 \right)
\nonumber\\
&& \hspace{-1cm}
 +\chi
 \left(
  \frac{\partial E_{\mathrm{q}}}
       {\partial n}
  \mbox{d} n
  +\frac{\partial E_{\mathrm{q}}}
   {\partial Q_{\mathrm{q}}}
   \mbox{d}Q_{\mathrm{q}}
 \right)
 +(E_{\mathrm{q}}-E_{\mathrm{N}})\mbox{d}\chi.
\label{dE01}
\end{eqnarray}
On the other hand, differentiating Eqs.~(\ref{rhotot}) and (\ref{Qtot})
at a given pair of $\rho_{\mathrm{t}}$ and $Q_{\mathrm{t}}$, we have
\begin{eqnarray}
(1-\chi)\mbox{d}\rho
&=& (\rho-n)\mbox{d}\chi-\chi\mbox{d} n,
 \label{drhoh}\\
(1-\chi)\mbox{d}Q_{\mathrm{N}}
&=& (Q_{\mathrm{N}}-Q_{\mathrm{q}})\mbox{d}\chi-\chi\mbox{d}Q_{\mathrm{q}}.
 \label{dQh}
\end{eqnarray}

To minimize $E_{\mathrm{t}}$, we substitute
Eqs.~(\ref{drhoh}) and (\ref{dQh}) into Eq.~(\ref{dE01}).
Then setting $\mbox{d}E_{\mathrm{t}}=0$, we find
\begin{equation} \label{Gibbscond}
\frac{\partial E_{\mathrm{N}}} {\partial \rho}
=\frac{\partial E_{\mathrm{q}}} {\partial n},\ \
\frac{\partial E_{\mathrm{N}}} {\partial Q_{\mathrm{N}}}
=\frac{\partial E_{\mathrm{q}}} {\partial Q_{\mathrm{q}}},\ \
P_{\mathrm{N}}=P_{\mathrm{q}},
\end{equation}
where
\begin{eqnarray}
P_{\mathrm{N}}
&=&
 -E_{\mathrm{N}}
 +\rho\frac{\partial E_{\mathrm{N}}}{\partial\rho}
 +Q_{\mathrm{N}}\frac{\partial E_{\mathrm{N}}}{\partial Q_{\mathrm{N}}},\\
P_{\mathrm{q}}
&=&
 -E_{\mathrm{q}}
 +n\frac{\partial E_{\mathrm{q}}}{\partial n}
 +Q_{\mathrm{q}}\frac{\partial E_{\mathrm{q}}}{\partial Q_{\mathrm{q}}}.
\end{eqnarray}
The conditions in Eq.~(\ref{Gibbscond}) are nothing but the Gibbs
ones, i.e., the baryon chemical potential, the charge chemical
potential, and the pressure in nuclear and quark matter should be
equal to each other to minimize the total energy of the mixed phase.

In the previous two sections, we have linked the baryon chemical potential and charge chemical
potential to the respective constituent particle chemical potentials in
Eqs.~(\ref{baryonchemmuud}) and (\ref{baryonchemNM}). As application of these equalities, we
immediately see that the first two equations in (\ref{Gibbscond}) are equivalent to
\begin{equation} \label{munpud}
\begin{array}{rcl}
\mu_{\mathrm{n}}&=&\mu_u+2\mu_d,\\
\mu_{\mathrm{p}}&=&2\mu_u+\mu_d.
\end{array}\ \
\mbox{or}\ \
\begin{array}{rcl}
\mu_u&=&(2\mu_{\mathrm{p}}-\mu_{\mathrm{n}})/3,\\
\mu_d&=&(2\mu_{\mathrm{n}}-\mu_{\mathrm{p}})/3.
\end{array}
\end{equation}

In general, all other chemical potentials in quark sector can be
related to $\mu_u$ and $\mu_d$, e.g., $\mu_s=\mu_d$,
$\mu_e=\mu_d-\mu_u$. Similarly, all chemical potentials in nuclear
sector can be linked to $\mu_{\mathrm{n}}$ and $\mu_{\mathrm{p}}$,
e.g., $\mu_e=\mu_n-\mu_p$. Therefore, Eq.~(\ref{munpud}) means that
we can choose either ($\mu_u$,$\mu_d$) or
($\mu_{\mathrm{n}}$,$\mu_{\mathrm{p}}$) as the two independent
chemical potentials. The latter can then be determined by solving
the charge neutrality equation and the pressure balance equation for
a given total baryon number or a given quark fraction.

At finite temperature, we similarly have the
phase equilibrium condition
\begin{eqnarray}
& P_{\mathrm{N}}=P_{\mathrm{q}} & \hspace{0.5cm} (\mbox{mechanical}),
 \label{phaseTcon01}\\
& \mu_{\mathrm{N}}=\mu_{\mathrm{q}} & \hspace{0.5cm} (\mbox{chemical}),
\label{phaseTcon02} \\
& T_{\mathrm{N}}=T_{\mathrm{q}}\equiv T  & \hspace{0.5cm} (\mbox{thermodynamical}).
\label{phaseTcon03}
\end{eqnarray}

The condition in Eq.~(\ref{phaseTcon03}) only tells us that
the temperature in nuclear and quark sectors are equal, so
we have a common temperature $T$. The chemical equilibrium
condition in Eq.~(\ref{phaseTcon02}) is equivalent to
that in Eq.~(\ref{munpud}). Therefore, we still have only
two independent chemical potentials. For a given total
density $\rho_{\mathrm{t}}$ at a fixed temperature $T$,
the two independent chemical potentials and the quark fraction $\chi$
can then be determined by solving the three equations in
Eqs.~(\ref{rhotot}), (\ref{Qtot}) with $Q_{\mathrm{t}}=0$,
and (\ref{phaseTcon01}).

Similar to the case at zero temperature, the lower critical density
$\rho_{\mathrm{c}1}$, which separates the nuclear and mixed phases,
is defined by $\chi=0$, while the critical density $\rho_{\mathrm{c}2}$
between the mixed and quark phases is determined by $\chi=1$.

\begin{figure}[htb]
\centering
\includegraphics[width=8.2cm]{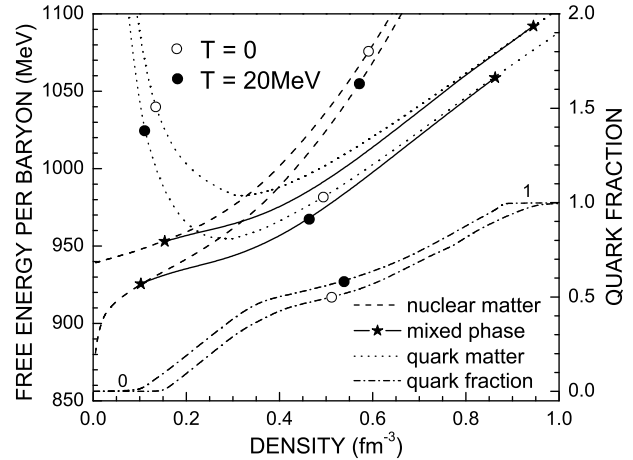}
\caption{
 The free energy per baryon as a function of density
 in the nuclear (dashed curves), mixed (solid),
 and quark phase (dotted). The quark fraction is also
 shown on the right axis with dash-dotted lines.
 The curves with an open circle are at zero temperature
 while those with a full circle are for the temperature $T=20$ MeV.
         }
 \label{mixE17095}
\end{figure}

\begin{figure}[htb]
\centering
\includegraphics[width=8.2cm]{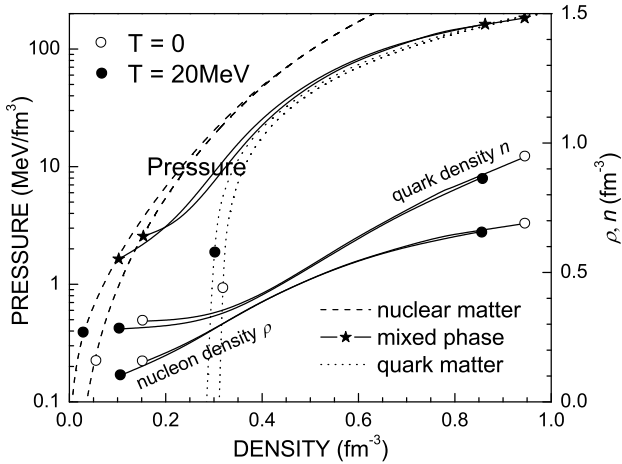}
\caption{ The pressure in nuclear (dashed curve), mixed (solid), and
quark phase (dotted) vs density at temperature $T=0$ (cures with an
open circle) and 20 MeV (cures with an solid circle). the nuclear
density $\rho$ and the quark density $n$ in the mixed phase are also
shown on the right axis.
         }
\label{mixP17095}
\end{figure}

In Fig.~\ref{mixE17095}, we display, for the parameter $\sqrt{D}$ =
170 MeV, the density dependence of the energy per baryon in pure
nuclear matter, in pure quark matter, and in the mixed phase. The
quark fraction has also been depicted on the right y-axis. In
Fig.~\ref{mixP17095} the corresponding pressure is reported. It is
seen that the nuclear matter is the most favorite phase at lower
densities, and the quark matter is the most stable phase at higher
densities, while at intermediate densities, the mixed phase has the
lowest energy.

The quark baryon number density $n$ and nuclear density $\rho$ are
also plotted on the right axis of Fig.~\ref{mixP17095}. We see that
the quark density is always higher than the nuclear density.
The transition from hadron phase to mixed phase occurs at the density a bit
less than $0.15$ fm$^{-3}$, well below the saturation density.
But it is hard to observe in terrestrial laboratories, since the
nuclear matter so far realized in exotic nuclei or heavy-ion collisions
is much less neutron rich. The transition from mixed phase to pure
quark phase occurs at the total density $\rho_{\mathrm{t}}=0.85$ fm$^{-3}$,
where the nuclear density is only $\rho=0.64$ fm$^{-3}$. The density
range of mixed phase is only slightly depending on the temperature,
at least in temperature range of interest in neutron stars.

\begin{figure}[htb]
\centering
\includegraphics[width=8.2cm]{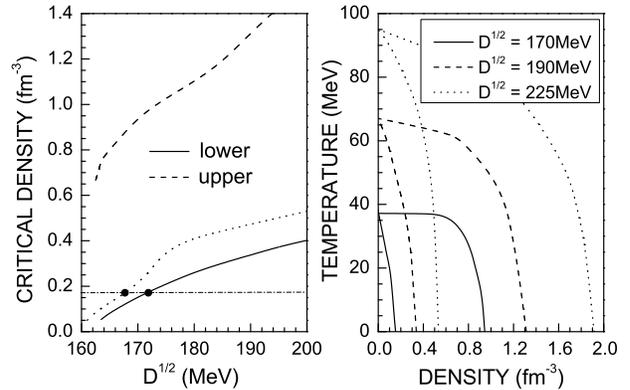}
\caption{ Left panel:
 The critical density of nuclear matter to quark matter
 as a function of the confinement parameter $D$. The horizontal line
 is the nuclear saturation density. The dotted line is the quark
 critical density in the case of two-flavor quark system.
Right panel:
 Phase diagram for three values of the confinement parameter
 and $m_{s0}=95$ MeV.
         }
 \label{rhocD}
\end{figure}

The critical densities depend on the parameter $D$. In the left
panel of Fig.~\ref{rhocD}, both nuclear critical density (the solid
line, separating the pure nuclear phase and the mixed phase) and the
quark critical density (the dashed line, delimiting the pure quark
phase) are displayed as a function of $D^{1/2}$. If $D^{1/2}<161.6$
MeV, the two critical densities approach zero, and accordingly SQM
is absolutely stable. When 161.6 MeV $<D^{1/2}<$ 162.5 MeV, mixed
phase can exist at any lower densities. Only when $D^{1/2}>162.5$
MeV, nuclear matter is more stable at lower densities. If we have
only two flavor quarks in the quark sector, the critical densities
are usually higher. In the same panel we also plot the lower
critical density for the two-flavor case. Because we know in our
real world the two-flavor quark matter does not exist below the
saturation density, $D^{1/2}$ should be on the right of the first
full dot at $D^{1/2}\approx 168$ MeV (the intersection of the dotted
and dot-dashed lines) in Fig.~\ref{rhocD}. On the other hand, to let
SQM have a chance to appear below the saturation density, $D^{1/2}$
should be on the left of the second full dot where $D^{1/2}= 171.3$
MeV. In plotting Fig.~\ref{mixE17095} and \ref{mixP17095}, we adopt
$D^{1/2}=170$ MeV. The temperature dependence of the lower and
higher critical densities is also plotted in Fig.~\ref{rhocD} (right
panel). The two lines at fixed $D$ mark the boundaries of the three
phases.

\section{Properties of hybrid stars}
\label{neutronstar}

With the equation of state that has the mixed and/or
quark phase derived in Sec.~\ref{phase},
we are ready to study the structure of hybrid stars by
solving the Tolman-Oppenheimer-Volkov equation
\begin{equation}
\frac{\mathrm{d}P}{\mathrm{d}r}
=-\frac{G m E}{r^2}
  \frac{(1+P/E)(1+4\pi r^3P/m)}{1-2Gm/r},
\end{equation}
where $G=6.707\times 10^{-45}$ MeV$^{-2}$ is the gravitational
constant, $r$ is the distance from the center of the star,
$E=E(r)$ and $P=P(r)$ are the energy density and pressure at the
radius $r$, respectively. The subsidiary condition is
\begin{equation}
\mathrm{d}m/\mathrm{d}r=4\pi r^2 E
\end{equation}
with $m=m(r)$ being the mass within the radius $r$.

At variance with pure nuclear or quark stars, a hybrid star contains pure
quark matter in the core, pure nuclear matter near the outer part, and,
in between, a mixed phase of the quark and nuclear matter. In this case,
therefore, we must use the EoS in whole density range.

\begin{figure}[htb]
\centering
\includegraphics[width=8.2cm]{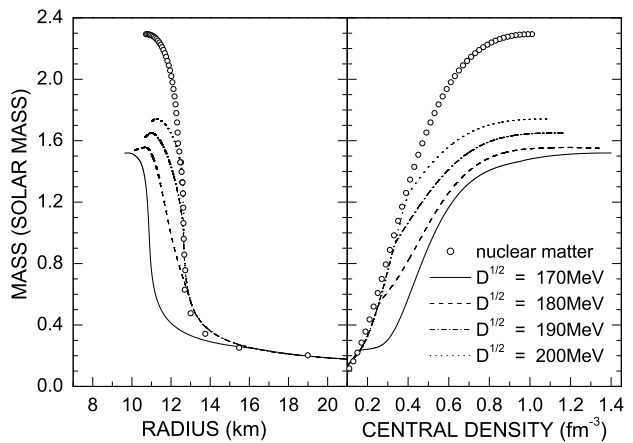}
\caption{
 Mass-radius relation of hybrid stars at $m_{s0}=95$ MeV for four values
 of the confinement parameter $D^{1/2}$ of 170, 180, 190, 200 MeV .
         }
 \label{mass}
\end{figure}

 The resulting gravitational mass for the hybrid star is plotted in
Fig.~\ref{mass}, as a function of both radius and central density,
for four values of the confinement parameter $D^{1/2}$ of 170, 180,
190, 200 MeV. The main effect of the phase transition is, as expected,
a large reduction of the maximum mass due to softening of the EoS.
Concerning the confinement parameter $D$, we observe a slight decrease
of the maximum mass when the $D$ value becomes smaller, from 1.74$M_\odot$
at $D^{1/2}$ = 200 MeV down to 1.6, 1.55, 1.52 $M_\odot$ corresponding
to $D^{1/2}$ = 190, 180, 170 MeV. This can be easily understood in this
way: since quark phase is occurring at lower density for smaller values
of $D$, sometimes even less than the nuclear saturation density (for
example only 0.15 fm$^{-3}$ for $D^{1/2}$ = 170 MeV), then the quark
population of the star would be more numerous, since the stronger
softening of the EoS may only support less gravitational mass. In
particular, the most favorable case of $D^{1/2}$ = 180 and 190 MeV,
when the quark confinement appears at around 2$\rho_0$ consistent
with the heavy ion experiment, the predicted maximum mass can match
very well the S branch of neutron stars mentioned by Haensel et al.\
with an error less than $5\%$ \cite{HAE}.

\begin{figure}[htb]
\centering
\includegraphics[width=8.2cm]{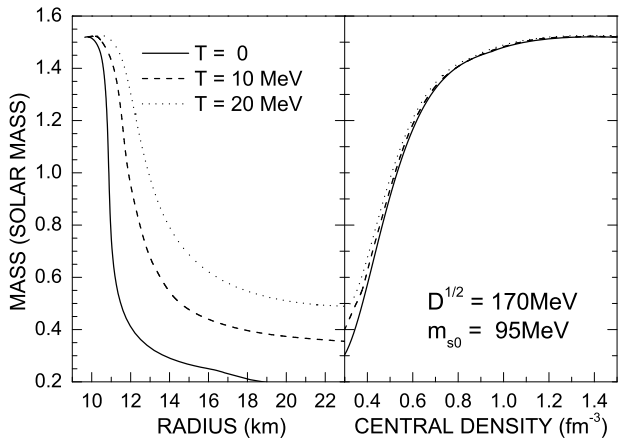}
\caption{
 Temperature effect on the mass-radius relation of hybrid stars.
        }
\label{mrT}
\end{figure}

In order to study the temperature effect on the mass of hybrid stars,
we plot the star mass, vs the star radius or central density, at $T=0$
and $T=20$ MeV in Fig.~\ref{mrT}. It is seen that the temperature
influence on the maximum mass is very limited. Otherwise the effect is
a quite strong increase of the NS radius for a fixed amount of gravitational
mass. But, larger is the mass smaller is the radius variation, similarly
to a few other calculations \cite{nico}.

\subsection{Kaon condensation}

The condensation of $K^-$ mesons in neutron stars is widely discussed
in the literature (see Refs.~\cite{kao1,Brown08} and references therein quoted).
In dense matter, the condensation of $K^-$ mesons is originated by the reaction
\begin{equation}
  e^- \rightarrow K^- + \nu.
\label{eq:kaon.1}
\end{equation}
if the effective mass of the $K^-$ drops below the chemical
potential of the electron, this reaction becomes possible in dense
matter, indicating the presence of kaon condensation. Since almost
all the studies of kaon medium properties \cite{kaon} suggest the
consistent picture that the attraction from nuclear matter would
bring the $K^-$ mass down, so the threshold condition for the onset
of $K^-$ condensation $\mu_e = m^*_K$, which follows from
Eq.~(\ref{eq:kaon.1}), could be fulfilled in the center
of neutron stars \cite{kaon} at $\rho \gtrsim 3\rho_0$. However, the
deconfinement phase transition from the hadronic phase to the quark
phase occurs also at rather low density, which leads to the onset
of mixed phase in neutron stars. Then it is very interesting to
explore how the quark deconfinement affects the $K^-$ condensation
threshold.

\begin{figure}[htb]
\centering
\includegraphics[width=8.2cm]{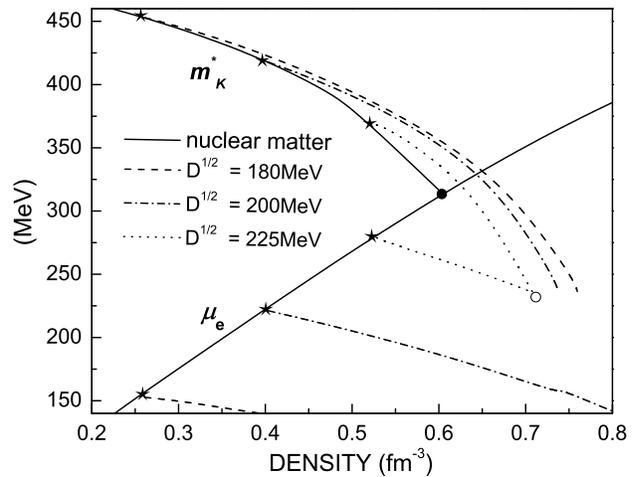}
\caption{ The effective kaon mass and the electron chemical
potential are shown in normal nuclear matter and also in mixed phase
at $m_{s0}=95$ MeV for three selected values of the confinement
parameter $D^{1/2}$ of 180, 200, 225 MeV.
         }
 \label{kaon}
\end{figure}

We take the antikaon dispersion relation constrained by the
heavy-ion data as empirical indication of an attractive antikaon
potential in dense matter \cite{kaon}, and combine this with the BHF
model of nuclear matter together with the above quark model without
bag constant, to calculate the effective kaon mass in hybrid stars.
The result is illustrated in Fig.~\ref{kaon}, where the effective
kaon mass and the electron chemical potential are shown in normal
nuclear matter and especially in mixed phase for three selected
values of the confinement parameter $D$. One sees that in normal
nuclear matter, $K^-$ medium mass decreases with increasing density and
meets the electron chemical potential at $\sim$ 0.6
fm$^{-3}$ (the solid bullet in Fig. 14), so the kaon
condensation would be present for sure in this case; However, once
the quark phase sets in at the total density of 0.26, 0.4, 0.53
fm$^{-3}$ respectively for $D^{1/2}$ = 180, 200, 225 MeV,
the decreasing speed of the kaon effective mass in the matter
slows down a little bit. In addition, more conclusively, the
electron population begin to decrease instead, for the reason that the
electric charge neutrality can be achieved more efficiently through
the charged quarks themselves. As a result the threshold condition
of $K^-$ condensation is much more difficult to satisfy, unless the
confinement parameter $D$ is chosen to be extremely high (at least
$D^{1/2}=225$ MeV) when the presence of quark phase is pushed to
very high densities, then one may expect that $m^*_K$ can finally
equal to $\mu_e$ (which is indicated with a circle). Those high
values of $D$ may even not be realistic, therefore
we would like to conclude that the inclusion of quark phase may
make the kaon condensation impossible in neutron stars, or at
least hinder it very strongly.

\section{Discussion and conclusions}

In this paper the neutron star inner structure has been studied.
The lack of strong observational constraints demands for
sophisticated models of the NS composition and interaction
mechanisms. In this study we only included baryons and quarks in
equilibrium with leptons and kaons. Since in the NS interior high
baryon densities are reached, N$\bar{\mathrm{N}}$ and nucleonic excitations
are expected to play a  major role. Their effects can be
incorporated in a three body force. The baryon EoS in weak coupling
equilibrium with electrons was derived within the BBG theory
suitably extended so to include the three body force. In the quark
sector, we adopted the semiphenomenological CDDM quark model, which
exhibits a confinement mechanism alternative to the crude MIT bag
model. Furthermore,  in contrast with the extension of the MIT
model, where the density dependence is introduced artificially \cite{ddbag},
CDDM quark model shapes its density dependence in agreement with chiral
requirements \cite{LombardoPRC72,PenggxNPA747}.

Hadrons and quarks in $\beta$-equilibrium were considered at zero
and finite temperature. Since typical temperatures of protoneutron
stars are as high as $40$ MeV and beyond, rigorous thermodynamic
potentials have to be derived. This was possible with CDDM quark
model, due to its simplicity, whereas some approximations are needed
to calculate the hadron phase in the Brueckner theory. The latter is
still a main drawback of the finite temperature microscopic theory
of strongly interacting Fermi systems.

The transition from the low density hadron phase to high density
quark phase in beta equilibrium was studied under the Glendenning
hypothesis of total charge neutrality. The Gibbs construction enabled
to follow the evolution of the mixed hadron-to-quark phase, varying
the temperature $T$ and the confinement parameter $D$. The EoS, in terms
of pressure, energy density and chemical composition, was
constructed as a function of $T$ and $D$. Moreover the phase diagram $T$-$D$
was also depicted. The transition density from hadron to mixed phase
is strongly dependent on the confinement parameter and, for D small
enough, it becomes lower than the nuclear saturation density.
However, at that point nuclear matter is so strongly neutron rich
that it hard to imagine that it can be observed in terrestrial
laboratory experiments. At high temperature the transition
density is comparatively smaller.

The TOV equations have been solved in the above discussed model of
neutron stars. The transition to quark matter produces a strong
reduction of the maximum mass, from $M=2.3M_\odot$ to $1.5M_{\odot}$
for the lowest value of the confinement parameter $D$, but the
corresponding radius is not changed, being in both cases about $10$
km. The other $M$-$R$ configurations ly in a range $R\simeq 10\sim 13$
km depending on the value of $D$. With increasing $D$, the maximum mass
increases from $1.5$ to $1.7$ solar masses. Other quark models have
been adopted to describe the NS transition to a deconfined phase
(see, for instance, Refs.~\cite{Schertler99prc,njl,nico}). Except for
the Nambu-Jona Lasinio model which exhibits instability of neutron
stars \cite{njl}, the other models point to a softening of the EoS
of nuclear matter. The MIT model with density-dependent bag constant
predicts a beta-stable quark phase quite similar to our model
\cite{nico}, but the comparison on the NS structure turns out to be
difficult, since in that calculation the transition to quark phase
is built on top of hyperonized nuclear matter.

The effect of temperature on the maximum mass is negligible up to
$T=20$ MeV, but as expected the heated system can support much less
gravitational mass in the same volume for all $M$-$R$ configurations.
Thus, an evolutionary study of isolated neutron stars would show a
strong compression from the newborn phase to the long era phase as
fast as the star cools down from $40$ MeV to approximately zero.

Beyond the confinement parameter, the transition to quark phase is
interrelated to other properties, in particular the possible onset
of kaon condensation. In our analysis we concluded that the quark
phase could not be compatible with the kaon condensation in neutron
stars, or at least could hinder it very strongly.

The model developed in this paper misses some important aspects.
The first one  is the effect of neutrinos. Neutrinos play the major
role in new born neutron stars, when they are still trapped in the
interior. In this case the neutronization process is still hindered
and the EoS is that of symmetric nuclear matter. Moreover, the onset
of kaon condensation is shifted to higher densities \cite{kao2} and
probably kaons have no more any chance to compete with quarks.

The second one is the inclusion of hyperons and their competition
with other mechanisms such as kaon condensation. In Refs.~\cite{nico}
the hybrid stars are studied by including hyperons in the hadron
phase, which makes the hadron EoS very soft and brings the maximum
mass to $M=1.25M_\odot$. The introduction of the quark phase rises
up again the maximum mass to $1.5M_\odot$, a value which is
consistent with our prediction without hyperons. On the other hand,
the appearance of hyperons much depends on the threshold of kaon
production.  
Therefore the interplay between hyperons and
kaons turns out to be quite important and deserves additional
investigation, as soon as more reliable empirical inputs will be
available, especially on the hyperon-nucleon and hyperon-hyperon
interaction.

\section*{Acknowledgments}
One of us (U.L.) is indebted to Dr M. Baldo for several discussions
on statistical thermodynamics.
G.X.\ would like to thank support from NSFC (10675137, 10375074, 90203004)
and KJCX2-YW-N2, and also acknowledge warm hospitality at
the Dipartimento di Fisica e Astronomia (Universit\'{a} di Catania),
the Laboratori Nazionali del Sud (INFN, Catania),
and the MIT Center for Theoretical Physics (MIT-CTP)..

\appendix

\section{Thermodynamics with confinement by the density dependence
of quark masses}
\label{appeA}

Let's start from the fundamental thermodynamic differentiation relation
$
\mbox{d}(VE)
=T\mbox{d}(VS)-P\mbox{d}V+\sum_i\mu_i\mbox{d}(Vn_i),
$
where $S$ is the entropy density, and (anti)particles are
assumed to be uniformly distributed in a volume $V$.
By using the free energy density $F=E-TS$, it becomes
$
\mbox{d}(VF)
=-VS\mbox{d}T-P\mbox{d}V+\sum_i\mu_i\mbox{d}(Vn_i),
$\
or, equivalently,
$
\mbox{d}F
=-S\mbox{d}T+(-P-F+\sum_i\mu_in_i)\mbox{d}V/V
 +\sum_i\mu_i\mbox{d}n_i.
$
Because of the uniformity, the free energy density
has nothing to do with the volume. We thus have
\begin{eqnarray}
P&=&-F+\sum_i\mu_in_i,
\label{Ptexpr} \\
\mbox{d}F&=&-S\mbox{d}T+\sum_i\mu_i\mbox{d}n_i.
\label{dFexpr}
\end{eqnarray}

 At finite temperature, we should consider both
particles and anti-particles, and particles/anti-particles
are not always located bellow the Fermi energy.
Therefore, 
The net particle number densities can be expressed as
\begin{equation}
n_i
=n_i(T,\mu_i^*,m_i)
=n_i^+-n_i^-.
\label{niTmum}
\end{equation}
where the superscripts $^+$\ indicates particles,
and $^-$ signifies antiparticles:
\begin{equation}
n_i^{\pm}
=g_i
 \int_0^{\infty}
 \frac{1}{1+e^{(\sqrt{p^2+m_i^2}\mp\mu_i^*)/T}}
 \frac{p^2\mathrm{d}p}{2\pi^2}.
\label{niTmuspm}
\end{equation}

In Eqs.~(\ref{niTmuspm}) and (\ref{niTmum}), $\mu_i^*$ are effective
chemical potentials of respective particles. If quark densities
are not density and/or temperature dependent, $\mu_i^*$ are
nothing but the actual chemical potentials. In our present case,
however, quark masses depend on both density and temperature,
in order to include the strong interaction between quarks.
Therefore, the real chemical potentials should be
derived according to fundamental thermodynamic laws.

Eq.~(\ref{niTmum}) gives, implicitly,
$\mu_i^*$ as a function of $T$, $n_i$, and $m_i$,
i.e.,
\begin{eqnarray}
\mu_i^*
 &=& \mu_i^*(T,n_i,m_i).
\label{muiTim}
\end{eqnarray}

To determine the thermodynamic properties,
we need to give a characteristic function.
At zero temperature, we use the energy density
in Eq.~(\ref{Emod}) due to zero entropy.
Now the temperature $T$ and the densities $n_i$
are chosen as the independent system variables,
we should, therefore, choose the free energy in Eq.~(\ref{FTtot})
as the characteristic function.

Differentiation of Eq.~(\ref{FTtot}) gives
\begin{eqnarray}
\mbox{d}F
&=&
 \sum_i
 \left[
  \left(
   \frac{\partial F_i}{\partial T}
   +\frac{\partial F_i}{\partial\mu_i^*}
    \frac{\partial\mu_i^*}{\partial T}
  \right) \mbox{d}T
  +\frac{\partial F_i}{\partial\mu_i^*}
   \frac{\partial\mu_i^*}{\partial n_i}
         \mbox{d}n_i
 \right.
\nonumber\\
&& \phantom{\sum_i[}
 \left.
  \left(
   \frac{\partial F_i}{\partial m_i}
   +\frac{\partial F_i}{\partial\mu_i^*}
    \frac{\partial\mu_i^*}{\partial m_i}
  \right)\mbox{d}m_i
 \right].
\end{eqnarray}

Applying
$
\mbox{d}m_i
=\frac{\partial m_i}{\partial T}\mbox{d}T
 +\sum_j\frac{\partial m_i}{\partial n_j}\mbox{d}n_j
$,
then comparing the corresponding expression
with Eq.~(\ref{dFexpr})
we immediately have
\begin{equation}
\mu_i
=
   \frac{\partial F_i}{\partial\mu_i^*}
   \frac{\partial\mu_i^*}{\partial n_i}
+\sum_j
 \left(
   \frac{\partial F_j}{\partial m_j}
   +\frac{\partial F_j}{\partial \mu_j^*}
    \frac{\partial\mu_j^*}{\partial m_j}
  \right)
  \frac{\partial m_j}{\partial n_i}.
\label{muimus}
\end{equation}
and
\begin{equation}
S=
 -\sum_i
  \left[
    \frac{\partial F_i}{\partial T}
   +\frac{\partial F_i}{\partial\mu_i^*}
     \frac{\partial\mu_i^*}{\partial T}
   +\left(
     \frac{\partial F_i}{\partial m_i}
     +\frac{\partial F_i}{\partial\mu_i^*}
      \frac{\partial\mu_i^*}{\partial m_i}
    \right)
    \frac{\partial m_i}{\partial T}
  \right]
\label{STmus}
\end{equation}

To simplify the expressions, we differentiate
Eqs.~(\ref{niTmum}) and (\ref{muiTim}) to get
\begin{eqnarray}
\mbox{d} n_i
&=&
   \frac{\partial n_i}{\partial T}\mbox{d}T
  +\frac{\partial n_i}{\partial \mu_i^*}\mbox{d}\mu_i^*
  +\frac{\partial n_i}{\partial m_i}\mbox{d}m_i, \\
\mbox{d}\mu_i^*
&=&
   \frac{\partial\mu_i^*}{\partial T}\mbox{d}T
  +\frac{\partial\mu_i^*}{\partial n_i}\mbox{d}n_i
  +\frac{\partial\mu_i^*}{\partial m_i}\mbox{d}m_i,
\end{eqnarray}
which implies
\begin{equation}
\frac{\partial\mu_i^*}{\partial T} \frac{\partial n_i}{\partial \mu_i^*}
= -\frac{\partial n_i}{\partial T}, \
\frac{\partial\mu_i^*}{\partial n_i} \frac{\partial n_i}{\partial \mu_i^*}
= -1, \
\frac{\partial\mu_i^*}{\partial m_i}\frac{\partial n_i}{\partial \mu_i^*}
=-\frac{\partial n_i}{\partial m_i}.
\label{dnidmui}
\end{equation}

Defining
\begin{equation}
\Omega_0\equiv\sum_i\Omega_{0,i}(T,\mu_i^*,m_i)
=\sum_i\left[\Omega_{0,i}^+
        +\Omega_{0,i}^-\right]
\end{equation}
with
\begin{equation}
\Omega_{0,i}^{\pm}
=-\frac{g_iT}{2\pi^2}
 \int_0^{\infty}
 \ln\left[1+e^{-(\sqrt{p^2+m_i^2}\mp\mu_i^*)/T}\right]
 p^2\mbox{d}p,
\end{equation}
then we can write $F_i=\Omega_{0,i}+\mu_i^*n_i$.
Substituting this into Eq.~(\ref{STmus}) and (\ref{muimus}),
then applying Eq.~(\ref{dnidmui}) and
$n_i=-\partial\Omega_{0,i}/\partial\mu_i^*$, we have
\begin{equation}
\mu_i
=\mu_i^*
 +\sum_j\frac{\partial\Omega_0}{\partial m_j}
        \frac{\partial m_j}{\partial n_i}.
\label{muiTmusexp}
\end{equation}
and
\begin{equation}
S=-\frac{\partial\Omega_0}{\partial T}
  -\sum_i\frac{\partial\Omega_0}{\partial m_i}
         \frac{\partial m_i}{\partial T}.
\label{STmusexp}
\end{equation}
Obviously, the free energy density can be given as
\begin{equation}
F=\Omega_0
  -\sum_i \mu_i^*
   \frac{\partial\Omega_0}{\partial\mu_i^*}.
\end{equation}
The energy density is obtained by $E=F+TS$, giving
\begin{equation}
E=\Omega_0
  -\sum_i\mu_i^*\frac{\partial\Omega_0}{\partial\mu_i^*}
  -T\frac{\partial\Omega_0}{\partial T}
  -T\sum_i\frac{\partial\Omega_0}{\partial m_i}
          \frac{\partial m_i}{\partial T}.
\end{equation}
And the pressure is obtained by substituting
Eq.~(\ref{muiTmusexp}) into Eq.~(\ref{Ptexpr}):
\begin{equation}
P
=-\Omega_0
 +\sum_{i,j}n_i\frac{\partial\Omega_0}{\partial m_j}
        \frac{\partial m_j}{\partial n_i}.
\label{PTmusexp}
\end{equation}
Eqs.~(\ref{STmusexp})-(\ref{PTmusexp}) here are in complete
accordance with  the Eqs.~(58)-(61) in \cite{WenxjPRC72}
if one regards the $\mu_i$ there
as $\mu_i^*$ and $\Omega$ as $\Omega_0$
which were not explicitly stated.
The expression in Eq.~(\ref{muiTmusexp}) is special
in the present paper because we need the real
chemical potential to investigate the mixed phase.

The real thermodynamic potential density of the system is
\begin{equation}
\Omega
=F-\sum_i\mu_in_i
=\Omega_0
 -\sum_{i,j}n_i\frac{\partial\Omega}{\partial m_j}
        \frac{\partial m_j}{\partial n_i}.
\end{equation}


In the above derivation, we choose volume $V$, the temperature$T$,
and the particle number densities $n_i$ as the independent system
variables. In this case, the free energy is the characteristic function
from which we get the complete set of thermodynamic functions.
For this purpose we have defined the intermediate variables $\mu_i^*$
in Eqs.\ (\ref{niTmuspm}) and (\ref{FiTmum}).
Because the quark matter we are considering is a strongly interacting
system, the relations between the chemical potentials and the densities
are, in principle, not the same as those of a free Fermi gas.
However, with the `effective' chemical potentials $\mu_i^*$,
the densities and the free energy are really of the same form
as those of a non-interacting Fermi gas. This is what the
`effective' means. The actual chemical potentials of each type
of particles are determined from the fundamental thermodynamic
equality (\ref{dFexpr}) which results in Eq.~(\ref{muiTmusexp}).















\end{document}